# Mode-matching in multiresonant plasmonic nanoantennas for enhanced second harmonic generation

*Michele Celebrano[1], Xiaofei Wu[2,3], Milena Baselli[1], Swen Großmann[2], Paolo Biagioni[1], Andrea Locatelli[4], Costantino De Angelis[4], Giulio Cerullo[1,5], Roberto Osellame[5], Bert Hecht[2], Lamberto Duò[1], Franco Ciccacci[1], & Marco Finazzi[1]*

[1]Physics Department, Politecnico Milano, P.zza Leonardo Da Vinci 32, 20133 Milano, Italy.

[2]Nano-Optics & Biophotonics Group - Department of Physics - Experimental Physics 5, University of Würzburg, Am Hubland, 97074 Würzburg, Germany.

[3]Present Address: Ultrafast Nanooptics Group – Department of Physics – Experimental Physics III, University of Bayreuth, Universitätsstraße 30, 95447 Bayreuth, Germany.

[4]Department of Information Engineering, University of Brescia, Via Branze 38, 25123 Brescia, Italy.

[5]IFN-CNR - Physics Department, Politecnico Milano, P.zza Leonardo Da Vinci 32, 20133 Milano, Italy.

ABSTRACT:

Boosting nonlinear frequency conversion in extremely confined volumes remains a key challenge in nano-optics, nanomedicine, photocatalysis, and background-free biosensing. To this aim, field enhancements in plasmonic nanostructures are often exploited to effectively compensate for the lack of phase-matching at the nanoscale. Second harmonic generation (SHG) is, however, strongly quenched by the high degree of symmetry in plasmonic materials at the atomic scale and in nanoantenna designs. Here, we devise a plasmonic nanoantenna lacking axial symmetry, which exhibits spatial and frequency mode overlap at both the excitation and the SHG wavelengths. The effective combination of these features in a single device allows obtaining unprecedented SHG conversion efficiency. Our results shed new light on the optimization of SHG at the nanoscale, paving the way to new classes of nanoscale coherent light sources and molecular sensing devices based on nonlinear plasmonic platforms.

The exploitation of low-dimensional structures such as metal films[1] and nanostructures[2] to boost nonlinear optical effects is becoming increasingly crucial in photonics,[3-5] given its implications on nonlinear frequency conversion at the nanoscale[6-9] that can be applied to all-optical signal processing as well as to life sciences: from local nonlinear phototherapy[10] to nonlinear sensing, photocatalysis,[11] tagging[12] and imaging.[13] SHG enhancement through plasmonic nanoantennas is often achieved by matching their localized surface plasmon resonance either with the excitation[14,15] or, more rarely, with the emission wavelength.[16] Double-gap nanoantenna designs featuring a doubly-resonant response at both the excitation and emission wavelengths[17] have also been proposed, while experimental SHG boosting has been achieved, to date, through the exploitation of extremely broad plasmonic resonances,[18] also in combination with Fano-like spectral features.[19] These features are often achieved by increasing the structure complexity, that eventually limits their scalability and/or increases the impact of local defects.

In this letter we report on the design and realization of a broadly-tunable plasmonic device based on coupled gold nanoantennas working in the near-infrared (NIR) (see Figure 1) and featuring multiple narrow plasmonic resonances. The investigated antennas exhibit unprecedented SHG efficiency at the nanoscale thanks to the fulfillment of specific criteria, namely (i) a multi-resonant response occurring at both the excitation and second harmonic (SH) wavelength, (ii) a significant spatial overlap of the localized fields at the wavelengths of interest and (iii) a geometry that fosters dipole-allowed SH emission.

First of all, a substantial improvement of SHG efficiency can be achieved by exploiting a multiresonant behavior with resonances simultaneously matching both the excitation and the SH wavelength.[17-19] In this doubly-resonant regime, SHG can be viewed as a coherent three-step process[20] (see Fig. 1a) in which the absorption of two photons at the fundamental wavelength (FW) assisted by a first plasmonic mode with energy $\hbar\omega$, is followed by a coherent radiative decay assisted by a second plasmonic mode oscillating exactly at twice the energy ($2\hbar\omega$), which restores the system ground state through the emission of a SH photon. The overall SHG rate depends on how efficiently the mode at $2\hbar\omega$ couples to the SH dipole excited by the electric field at $\hbar\omega$ through the nonlinear antenna polarizability. Therefore, to improve the efficiency of the process, both plasmonic modes need to be engineered to optimize such a coupling, which is also subject to parity and angular momentum conservation rules.[20-22] Ultimately, efficient SHG thus requires that the SH dipoles generated by the FW field efficiently overlap with the current distribution inside the material associated with the mode responsible for SH emission. Such "mode matching", which in non-centrosymmetric bulk materials corresponds to the well-known phase matching conditions, is essential to ensure efficient nonlinear frequency conversion at the nanoscale. In practice, for plasmonic antennas, a very good degree of mode overlap is already achieved when the involved modes generate field enhancement in the same nanoscale volume.

Moreover, light-matter interaction at the nanoscale follows a well-known hierarchy, depending on the multipole expansion of the electromagnetic field: in the long-wavelength limit $\lambda \gg a$, $a$ being the particle size and $\lambda$ the light wavelength, the transition probability rapidly falls off with increasing order of the (electric or magnetic) multipole associated with the transition, with electric multipoles giving higher transition rates than magnetic multipoles of the same order. For this reason, the optimization of SHG in sub-wavelength particles necessarily requires that all the three transitions indicated in Fig. 1a should be electric-dipole-allowed. This condition cannot be satisfied in particles characterized by either axial or inversion symmetry, since an electric dipole transition couples states with opposite parity. To this aim, one or both modes of the structure need to simultaneously display a pronounced electric dipole (odd) and electric quadrupole (or magnetic dipole, both even) character.

Our nanostructure design is developed using Finite Difference Time Domain (FDTD) simulations (FDTD Solutions v 8.9, Lumerical Solutions, Inc., Canada). The geometry of each antenna, as inferred from high-resolution SEM images, is discretized by using a mesh with a 2-nm step size. Perfectly-matched absorbing boundary conditions are employed, while the dielectric constant of Au is taken from Ref. 23. The devices consists of two antenna elements closely coupled via a very small gap (see Fig. 1b): a V-shaped nanoantenna featuring multiple plasmonic resonances[24] and a single nanorod. The isolated V-shaped antenna displays two main modes, $V_1$ and $V_2$ (see dashed line in Fig. 1c), that can be tailored by varying either the angle between the arms, their thickness, and/or their width.[25] In order to both achieve optimal frequency overlap and maximize the mode-matching for the excitation and the SHG wavelengths, we increase the number of available degrees of freedom by finalizing the device by means of a rod-shaped antenna that is coupled to one of the V-shaped antenna arms through a gap of about 17 nm (see inset in Fig. 1b). At variance with the recent implementations of rod-shaped antennas as passive elements for SHG enhancement in nanoparticle ensembles,[26] here the nanorod directly couples with the active structure. As a consequence, by tuning the nanorod first-order longitudinal mode to match the SH emission line, hybridization with the $V_2$ mode is achieved, yielding a bonding mode, $V_2^B$, and an anti-bonding mode, $V_2^A$ (see solid line in Fig. 1c). As it will be shown later, the parameters of the devised nanoantenna geometry can be easily tailored to exactly match mode $V_2^A$ with the expected SHG line without substantially affecting $V_1$. Hence, in the finalized device $V_1$ still overlaps with the excitation laser frequency ($\omega$) to enhance absorption, whereas $V_2^A$ is tuned at $2\omega$ to boost the emission process, as illustrated in Fig. 1c. Concurrently, an excellent spatial mode overlap is achieved on the structure, as shown by the field enhancement maps in Figure 1d-f. The combination of field and charge distributions also indicates that, while mode $V_2^A$ displays both a quadrupolar behavior and a strong - horizontal (along the *x*-axis in Fig. 1d) - dipolar emission, mode $V_1$ is best excited by a vertical linear polarization (along the *y*-axis in Fig. 1f). This is ascribed to the properties of $V_2$ that are transferred by the hybridization and, as explained above, constitutes a crucial feature to enable dipole-allowed coupling of the $V_2^A$ mode with both the ground state and the electric dipole associated with the $V_1$ mode.

We engineered the device for excitation with NIR light at ~ 1560 nm wavelength to achieve SHG devices with emission ~ 780 nm, hence in a region where the absorption due to the interband transitions in gold is extremely weak. This key-enabling feature, employed only in a limited number of experimental approaches so far,[18] allows minimizing the SHG re-absorption by the metal and may become crucial in label-free biomedical imaging, given the low NIR absorption of biological tissues at these wavelengths.

Nanoantennas are fabricated from a single-crystalline gold flake via focused ion beam (FIB) milling[27], which allows realizing narrow and reproducible gap sizes (see the inset of Fig. 1b). The fabrication process is reported in Ref. 28. The structure width is designed to be ~ 30 nm while the thickness is set by the flake height (~ 40 nm). The ability to realize highly-reproducible structures with uncertainty below 10 nm is crucial since it minimizes spurious SHG from defects[14,29,30] that hinder systematic SHG efficiency optimization.

To validate our concept we realize a 6×6 array of nanostructures in which the relevant geometrical parameters are systematically varied such that a doubly-resonant antenna is expected to appear at the center of the array (see the SEM image in Fig. 2a). We employed dark-field spectroscopy to collect linear scattering spectra from each individual antenna. Figure 2b shows the superposition between the simulated and the dark-field scattering spectra obtained for the doubly-resonant antenna, emphasizing the high level of nanoscale geometrical control achieved with nanofabrication. The excellent correlation between design and realization is further evidenced through the comparison of contour plots featuring dark-field and calculated scattering spectra acquired on the two most significant array lines that cross at the doubly-resonant particle (see Fig. 3a-b and red and green frames in Fig. 2a, respectively). The analysis is limited to the visible-NIR region due to the spectrometer sensitivity range, however no significant deviations were measured in the NIR through confocal microscopy maps at 1550 nm (not shown). The tunability of this device is also demonstrated since, by either varying the rod length (Fig. 3a) or the V-shape size (Fig. 3b), the plasmon resonances of interest shift while the others are only slightly perturbed.

We excited the nanoantennas using ultrashort pulses (70-fs, ~ 1560-nm wavelength, 80-MHz repetition rate) from an amplified Er:fiber laser (Toptica Photonics AG) coupled to a 1.35 NA oil-immersion objective. The pulse width on the sample is estimated to be around 120 fs due to dispersion accumulated in the optical path and the laser power at the sample is set to 50 µW, which is low enough to exclude any photodamage. The signal is collected through the objective in epi-reflection geometry and sent to the detection path via a beam-splitter. To record the intensity maps we filter a region of about 40 nm around 780 nm using a narrow band-pass filter; the light is then sent to a single photon avalanche photodiode. The sample is mounted on a piezoelectric stage (Physik Instrumente GmbH & Co.) which is raster scanned. The signal on each image pixel is integrated for 50 ms and the pixel-size is set to 150 nm.

Figure 4a displays a confocal map of the SHG from the nanostructure array. Sizable SHG is obtained for a linear polarization that is vertically aligned to match the $V_1$ mode and higher emission occurs for V-shape arm lengths whose resonances lie close to the excitation laser spectrum. This holds true both for isolated V-shapes (see SEM image in Fig. 4b) and for coupled structures (right and left panels of Fig. 4a, respectively). However, when the SH intensity collected from the array of isolated V-shapes is compared with the one of the coupled structures, a strong intensity modulation induced by the rod presence can be observed. SHG reaches its highest value for rods with a resonance matching the SH wavelength, something that cannot be merely attributed to a superposition of the emissions from the individual nanostructures, since the SHG signal from the isolated nanorods is undetectable in our experimental conditions (not shown).

We modeled our experimental results by calculating the SHG emitted by the nanostructures using frequency-domain Finite Element Methods (FEM). The SH emitted by the nanostructure is numerically simulated using a frequency-domain finite-element solver (Comsol Multiphysics, Comsol, Inc., USA). Simulations are run following a perturbative approach in the so called undepleted-pump approximation (i.e. assuming that the SHG field does not couple back to the excitation field), and considering only the dominant contribution due to the free-electron currents normal to the metal surface.[31,32] The SHG map is created by taking the 6 × 6 array of SHG theoretical values obtained for the each nanostructure in the array and convolving it with a Gaussian 2D function with full-width at half-maximum defined by the resolution of our setup for SHG (~750 nm). The obtained theoretical 2D SHG map is reported in Fig. 4c and shows a very good agreement with the experimental map in Fig. 4a. In particular, both the experimental and the theoretical map shows a pronounced SHG from the same nanostructure, corresponding to the doubly-resonant antenna (indicated by the white arrows in the maps) optimized through FDTD simulations (see Fig. 1 and 3). These results demonstrate the effectiveness of our approach to deterministically optimize plasmon-enhanced SHG.

A full polarization analysis of the SH intensity has been performed on the doubly-resonant structure, and is reported in the angular plot of Fig. 4d. As expected from SHG selection rules using strongly focused beams[21], the emission behavior is the one of the electric dipole associated with mode $V_2^A$. This result also reveals a behavior similar to *Type I* phase matching in bulk materials, where the SH is emitted with a polarization perpendicular to that of the FW.

By combining a short-pass filter and a high-sensitivity spectrometer, we acquired the entire nonlinear emission spectrum of the doubly-resonant particle from 400 to 790 nm. The emitted spectra are integrated for about 30 s to compensate for the signal losses due to fiber coupling. Analysis of the spectrum reported in Fig. 4e reveals that in our devices Third Harmonic Generation (THG), which in nanostructured systems is often by far the dominating nonlinear process,[6] displays an intensity comparable to that of SHG. These nanostructures also feature an extremely low two-photon photoluminescence yield in this operational wavelength range, ensuring the emission of an almost pure coherent radiation. A comparison between the excitation laser spectrum and the SHG one (red and dark blue line in Fig. 4e inset, respectively) shows that, while the SH lies exactly at twice the energy of the FW, as expected, its bandwidth is slightly narrower than the theoretical value estimated by self-convoluting the laser spectrum (light blue line). This result further suggests that the SH emission is mediated by the nanoantenna plasmonic resonances.

Once the losses in the detection path of the setup are characterized and accounted for, we estimate that up to $3 \times 10^6$ photons/s are emitted by the doubly-resonant device, corresponding to a nonlinear coefficient $\gamma_{SHG} = \frac{P_{SHG}}{P_{FW}^2} \cong 5 \times 10^{-5}$ W$^{-1}$, $P_{SHG}$ being the SHG detected power and $P_{FW}$ the average incident power. This value remains constant while varying the incident power over one order of magnitude, further indicating the absence of photodamage. The reported value of $\gamma_{SHG}$ is almost 2 orders of magnitude larger than the one measured in the same experimental conditions from 100-nm-diameter gold spheres, used as reference structures (not shown), thanks to the extensive nonlinear optical characterization available in the literature[15,21,22,31-34]. Recently reported plasmonic structures featuring either a broadband resonance[18] or a broken-symmetry geometry,[35] investigated under similar experimental conditions, demonstrated efficiencies that are, at best, about one order of magnitude lower than ours. This confirms once again that the effective combination of a broken axial symmetry together with multiple plasmonic resonances and spatial mode-overlap can significantly boost the SHG process at the nanoscale.

In summary, we conceived and fully-characterized a plasmonic device based on a non-centrosymmetric gap antenna that features a by far unprecedented SHG efficiency. This is achieved through the effective combination on the same nanostructure of a plasmonic multiresonant character, spatial overlap of the plasmonic modes involved in the process and a broken-symmetry geometry. Our results also represent the first experimental proof of highly efficient SHG obtained from single-crystal and ultra-smooth nanostructures. The high degree of coherence of the light emitted, in the reported wavelength range, will allow realizing high-purity sub-wavelength nonlinear coherent sources that can be applied to achieve coherent control of light-nanostructure interaction mediated by SHG.[36] Furthermore, structures realized using our approach may be used as building blocks in label-free biomedical imaging, given the low absorption of biological tissues at both FW and SH wavelengths. Moreover, this concept holds promise for enhancing other second order nonlinear processes, i.e. parametric down conversion and difference frequency generation, offering new paths in nanoscale quantum-optics through the implementation of plasmonic nonlinear logic elements.

## Acknowledgements

The authors would like to thank Dr. Vikas Kumar and Tommaso Zandrini for valuable help in setting up the experiment and Dr. Riccardo Sapienza for providing the spherical nanoparticle sample. This work was funded by Fondazione Cariplo through the project SHAPES (id: 2013-0736). GC acknowledges support by the EC through the Graphene Flagship project (contract no. CNECT-ICT-604391). This work was performed in the context of the European COST Action MP1302 Nanospectroscopy

**Figure 1**

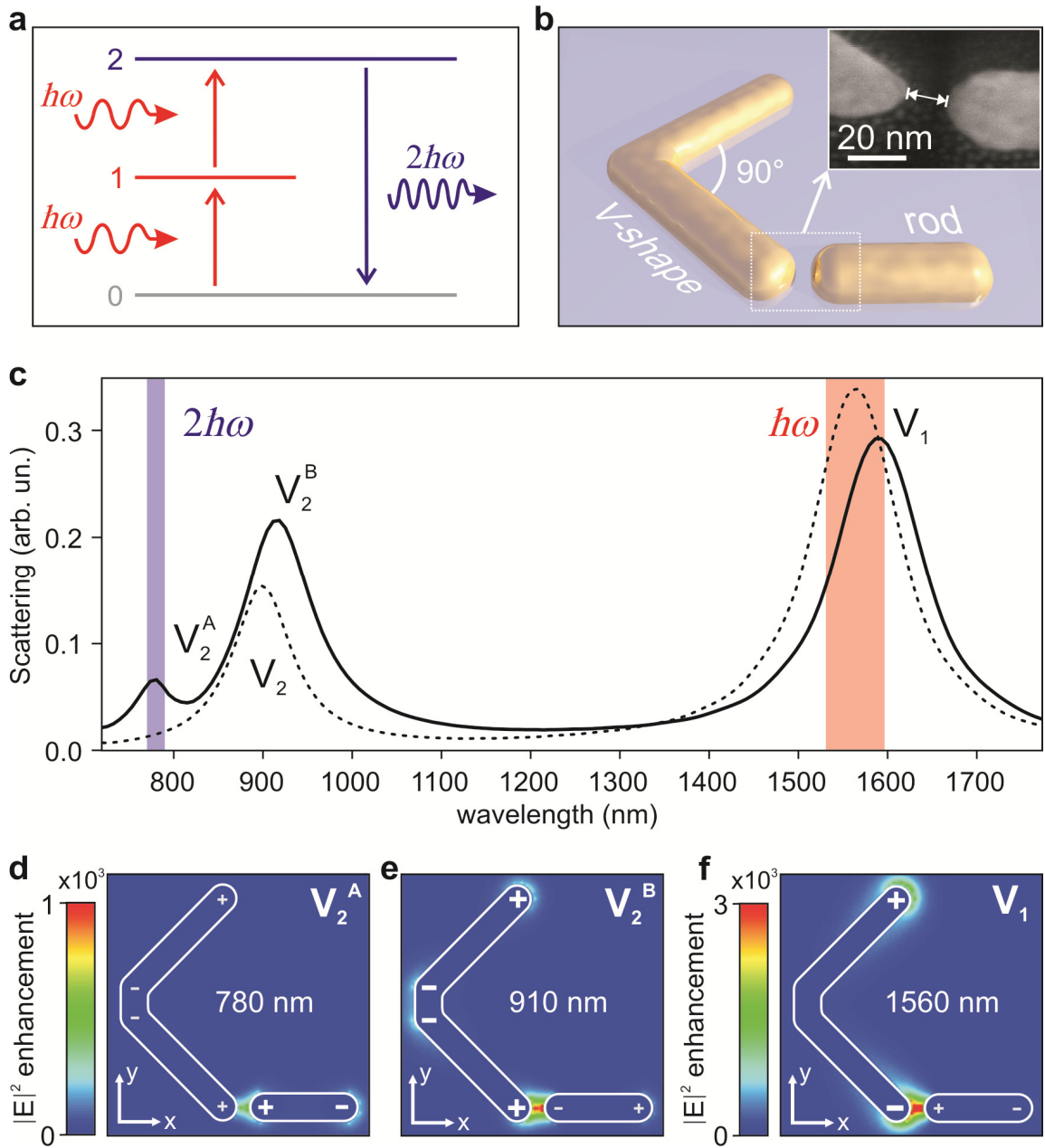

**Figure 1.** a) Scheme of the fundamental dipole transitions involved in the multiresonant plasmon-induced SHG process. b) Sketch of the engineered nanostructure for SHG enhancement. Inset: SEM image of the gap-region revealing a gap-size of about 17 nm. c) Scattering spectra of the isolated V-shape antenna (black dashed line) and of the coupled structure (black solid line), calculated by FDTD using unpolarized incident light. The black solid line also indicates that the coupled particle represents a finalized device, since its main modes, $V_1$ and $V_2^A$, do overlap with the excitation laser (light red stripe) and the expected SH (light blue stripe) bands, respectively. d-f) Local field and charge distributions relative to the structure main resonances. The dipolar modes are oriented along $x$ in (d) and (e) and along $y$ in (f).

**Figure 2**

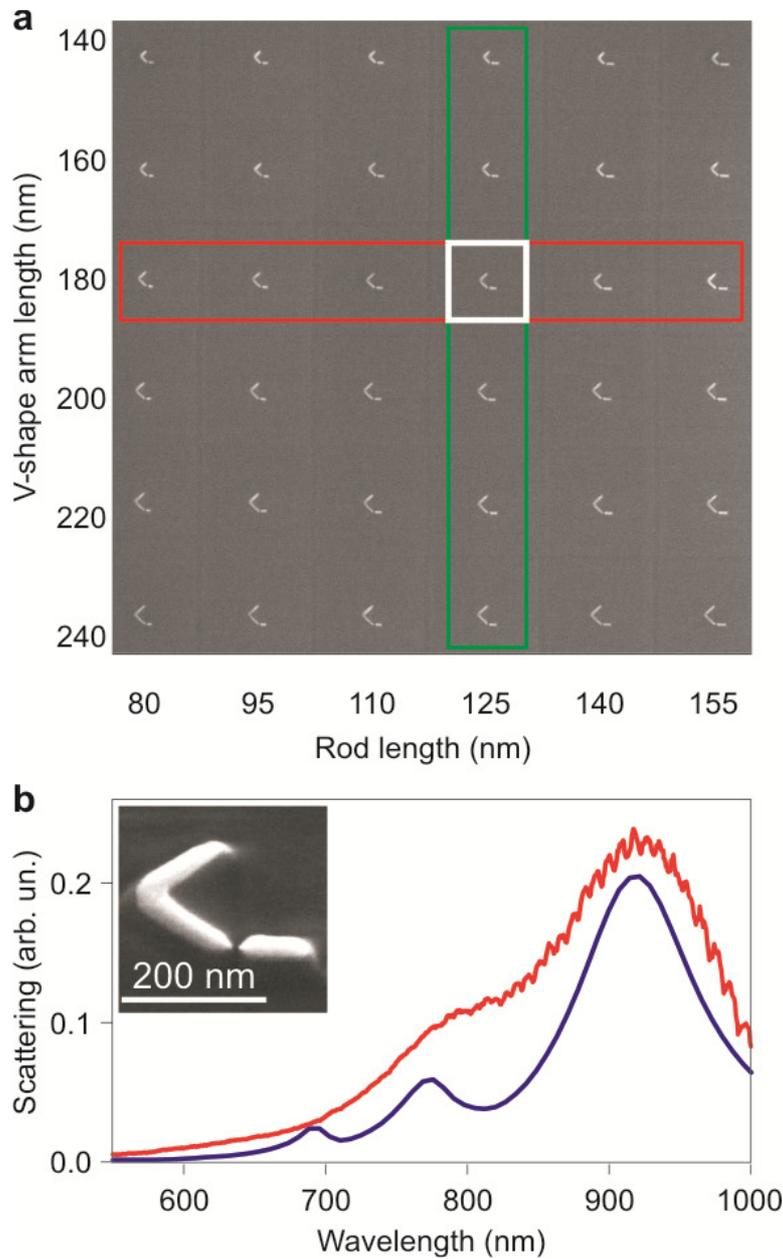

**Figure 2.** a) SEM image of the 6×6 array of nanostructures under investigation. The rod length increases from 80 nm to 155 nm in steps of 15 nm from left to right, while the V-shape half-arm length varies from 140 nm to 240 nm in steps of 20 nm from top to bottom. The white square indicates the expected doubly-resonant particle. The red rectangle comprises the row in which the rod length varies whereas the V-shape maintains an optimized length, while the green rectangle indicates the column in which the V-shape single arm changes lengths while the rod maintain an optimized length. b) Experimental scattering spectrum of the doubly-resonant nanostructure (red line) and its simulated counterpart (blue line). Inset: SEM image of the doubly-resonant nanostructure in the white square of panel (a). The additional feature appearing at 690 nm in the calculated spectrum (blue line) refers to a higher-order mode in the V-shape antenna.

**Figure 3**

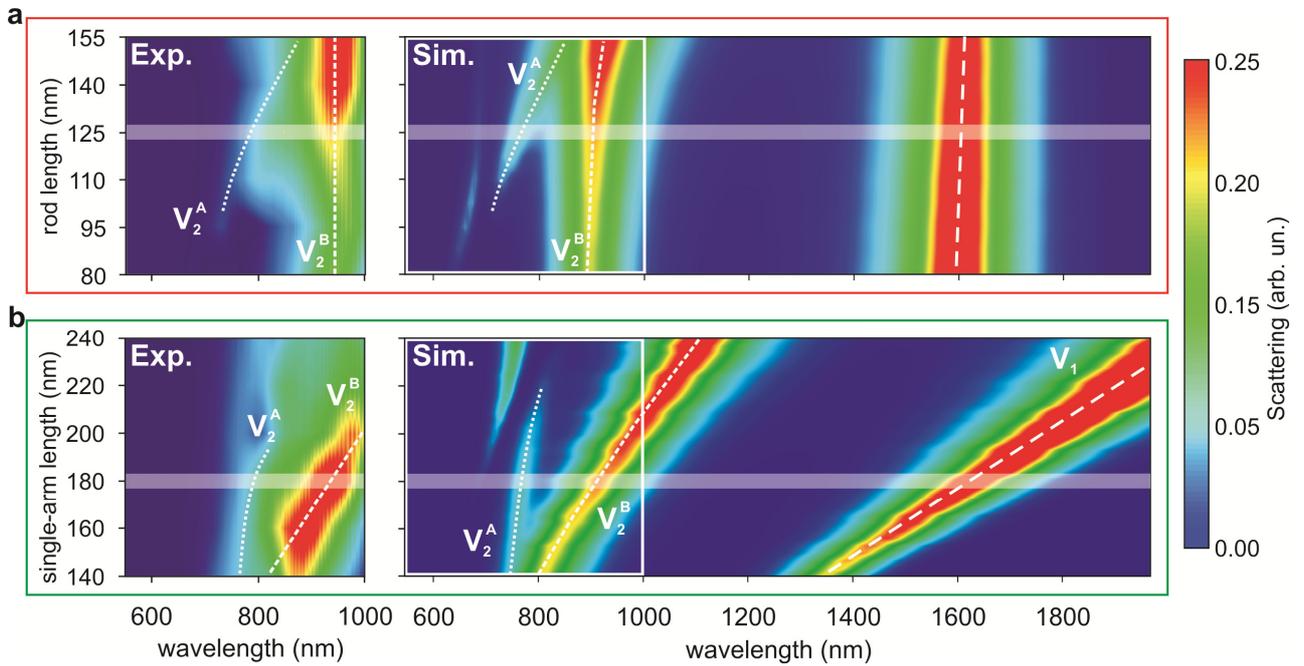

**Figure 3. Single nanostructures scattering spectra as a function of the antenna geometry.** a) Contour plots of the scattering spectra experimentally acquired using dark-field spectroscopy (left) and calculated using FDTD (right) in the visible-NIR on the nanostructures in the row indicated by the red rectangle in Fig 2a. b) Contour plots of the scattering spectra experimentally acquired using dark-field spectroscopy (left) and calculated using FDTD (right) in the visible-NIR on the nanostructures in the column indicated by the green rectangle in Fig 2a. The white lines are guides to the eye, to help identifying the different modes: $V_1$ (long-dashed line), $V_2^B$ (medium-dashed line) and $V_2^A$ (short-dashed line). The additional feature around 700 nm wavelength in the simulated spectra refers to a higher-order mode of the V-shape (see also Fig. 2b). The red and green frames help identifying the relative regions in the SEM map in Fig. 2a. The white semi-transparent bands indicate the spectrum of the doubly-resonant antenna.

**Figure 4**

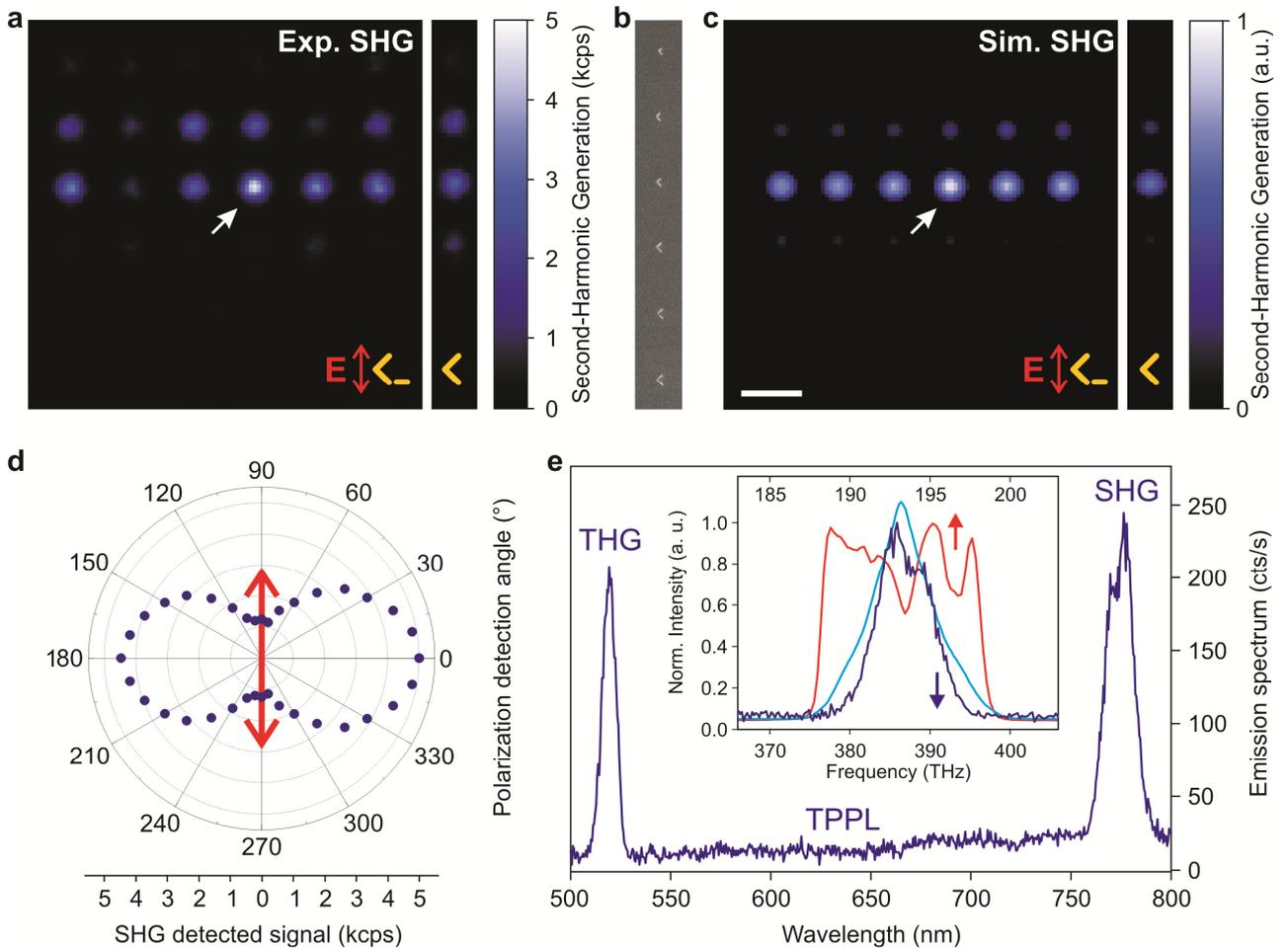

**Figure 4.** a) Left panel: SHG map collected from the array of nanostructures displayed in Fig. 2a. The double-headed red arrow indicates the impinging light polarization. Right panel: SHG collected from isolated (no coupled rod) V-shape structures with arm length varying from 140 nm to 240 nm (top to bottom) in 20-nm steps after excitation with the same polarization as in the left panel. b) SEM image of the isolated V-shape structures. c) Left panel: map of the simulated SHG from the same array with light polarization as in (a). Right panel: map of the simulated SHG from the array of isolated V-shapes presented in (b). d) The experimental polar plot (mirrored top-down) for the SHG collected from the resonant nanostructure (see white arrow in (a) and (c)). The double-headed red arrow indicates the impinging light polarization. e) Visible-NIR spectrum of the light emitted by the doubly-resonant device. The THG peak is centered around 519 nm, while the SHG one is centered at 776 nm. Inset: overlap between the SHG peak (dark-blue line) and the excitation laser band (red line). The theoretical SHG band obtained by self-convoluting the laser spectrum is also sketched (light-blue line). Horizontal scales are expressed in Hz and the experimental and theoretical SHG peaks (FWHM $\cong$ 7.8 THz and 10.3 THz respectively) are plotted on a frequency scale which is double the scale of the laser peak (FWHM $\cong$ 10.5 THz). In all measurements the excitation power is set to 50 µW.